\documentclass[12pt]{article}
\pdfoutput=1
\usepackage{graphicx}
\usepackage{epsfig}
\usepackage{epstopdf}
\DeclareGraphicsExtensions{.pdf,.eps,.png,.jpg,.mps}
\usepackage{caption}
\usepackage{float}
\usepackage{color}

\usepackage{url}
\usepackage[implicit=false]{hyperref}
\usepackage{cite}

\def\lsim{\mathrel{\rlap{\lower3pt\hbox{\hskip0pt$\sim$}}
     \raise1pt\hbox{$<$}}}         
\def\gsim{\mathrel{\rlap{\lower4pt\hbox{\hskip1pt$\sim$}}
     \raise1pt\hbox{$>$}}}         

\usepackage{amsmath}
\usepackage{amsfonts}

\begin{document}
\begin{titlepage}

\centerline{\Large \bf Non-Stationary Saturation of Inhomogeneously}
\centerline{\Large \bf Broadened EPR Lines}
\medskip

\centerline{Zura Kakushadze$^\S$$^\dag$\footnote{\, Zura Kakushadze, Ph.D., is the President of Quantigic$^\circledR$ Solutions LLC,
and a Full Professor at Free University of Tbilisi. Email: \href{mailto:zura@quantigic.com}{zura@quantigic.com}}}
\bigskip

\centerline{\em $^\S$ Quantigic$^\circledR$ Solutions LLC}
\centerline{\em 680 E Main St \#543, Stamford, CT 06901\,\,\footnote{\, DISCLAIMER: This address is used by the corresponding author for no
purpose other than to indicate his professional affiliation as is customary in
publications. In particular, the contents of this paper
are not intended as an investment, legal, tax or any other such advice,
and in no way represent views of Quantigic$^\circledR$ Solutions LLC,
the website \url{www.quantigic.com} or any of their other affiliates.
}}
\centerline{\em $^\dag$ Free University of Tbilisi, Business School \& School of Physics}
\centerline{\em 240, David Agmashenebeli Alley, Tbilisi, 0159, Georgia}
\medskip

\centerline{(September 3, 1990; in LaTeX form: January 17, 2020)\footnote{\, This note in Russian was published in 1991 in \cite{TSU}. I worked on this project while still in high school. For reasons outside my control, it was a few years before it was submitted to the journal. This English translation closely follows the original Russian version, with minor changes such as equation formatting and some additional references and explanatory footnotes.
}}

\bigskip
\medskip

\begin{abstract}
{}Non-stationary saturation of inhomogeneously broadened EPR lines is studied when cross-relaxation has the characteristics of spectral diffusion. A system of generalized kinetic equations is solved in quadratures in this approximation. The result is valid not only when the contribution of the spectral diffusion is negligible or dominant, but also in the intermediate case.    
\end{abstract}
\medskip

\end{titlepage}

\newpage
\section{Introduction}

{}Experimentally observed EPR (electron paramagnetic resonance) lines usually are broadened inhomogeneously \cite{Portis} and are described by the inverse temperatures $\beta(\omega, t)$ and $\beta_d(t)$ of the spin packet (SP) with the frequency $\omega$ and the dipole reservoir (DR), respectively \cite{BZK1}.\footnote{\, For additional related literature, see, e.g., \cite{BZK2}, \cite{BBZ}, \cite{BZK3}, and references therein.} Saturation of such systems has been studied in detail in the stationary case ($t\rightarrow\infty$). In this note we study non-stationary saturation of inhomogeneously broadened EPR lines, whose dynamics is described by the following system of generalized kinetic equations \cite{BZK1}, \cite{Thesis}:
\begin{eqnarray}
 &&\partial \beta(\omega, t) + {{\beta(\omega, t) - \beta_L}\over T_{SL}} + \pi\omega_1^2\varphi(\omega - \Omega)\left[\beta(\omega, t) + {{\Omega - \omega} \over \omega} \beta_d(t)\right]-\nonumber\\
 &&~~~~~~~ -{1 \over \omega}\int d\omega^{\prime}~g(\omega^{\prime} - \omega_0)~W_{CR}(\omega^{\prime} - \omega)\times \nonumber\\
 &&~~~~~~~\times \left[\omega^{\prime} \beta(\omega^{\prime}, t) - \omega \beta(\omega, t) + (\omega - \omega^{\prime})\beta_d(t)\right] = 0\label{Eq1}\\
 &&\partial\beta_d(t) + {{\beta_d(t) - \beta_L}\over T_{DL}} + {1 \over \omega_d^2}\int d\omega~g(\omega - \omega_0)~\omega~(\omega - \Omega)\times\nonumber\\
 &&~~~~~~~\times \left[\partial \beta(\omega, t) + {{\beta(\omega, t) - \beta_L}\over T_{SL}}\right] = 0\label{Eq2}
\end{eqnarray}
Here: $\partial$ denotes the time derivative; $\beta_L = \mbox{const}$ is the inverse temperature of the lattice; $\omega_1$ and $\Omega$ are the semi-amplitude and the frequency of the UHF (ultrahigh frequency) field; $\omega_0$ is the Zeeman frequency of the external constant magnetic field; $T_{SL}$ and $T_{DL}$ are the SP and DR spin-lattice relaxation times, respectively; $\omega_d$ is the DR energy ``quantum"; $W_{CR}(\omega^{\prime} - \omega)$ is the probability of cross-relaxation (CR); $\varphi(x)$ and $g(x)$ are the homogeneous and inhomogeneous line forms, respectively.

\section{Notations and Approximations}

{}Let $\Delta^*$, $\Delta_1$ and $\Delta_{CR}$ be the widths of the inhomogeneous line form, the hole burned therein, and the CR line form, respectively. Usually the following condition holds \cite{BZK4}:
\begin{equation}\label{hole}
 \Delta_1 \ll \Delta^* \ll \omega_0,\Omega
\end{equation}
That is, a narrow hole is burned in the inhomogeneously broadened line, which we assume hereinafter. We will also assume that\footnote{\, Non-stationary saturation of inhomogeneously broadened EPR lines under effective CR, where we have $\Delta_{CR} \gg \Delta^*$, was studied in \cite{Bulletin}.}
\begin{equation}\label{CR}
 \Delta_{CR} \ll \Delta^*
\end{equation}
With (\ref{hole}) and (\ref{CR}), the system (\ref{Eq1}) and (\ref{Eq2}) can be {\em approximated} as follows:
\newpage
\begin{eqnarray}
 &&\int d\omega^\prime~f(\omega^\prime - \omega)\left[\gamma(\omega^\prime,p) - \gamma(\omega, p)\right] -\nonumber\\
 &&~~~~~~~-\left[p + 1 + \pi\omega_1^2 ~T_{SL}~\varphi(\omega-\Omega)\right]\gamma(\omega,p) = \pi\omega_1^2~T_{SL}~\varphi(\omega-\Omega)\label{gamma1}\\
 &&\left[p + {T_{SL} \over T_{DL}(p)}\right]\gamma_d(p) + {(p+1)~\Omega \over \omega_d^2} \int d\omega~g(\omega - \omega_0)~(\omega - \Omega)~\gamma(\omega, p) = 0\label{gamma2}
\end{eqnarray}
Here
\begin{eqnarray}
 &&f(x) = T_{SL}~g(\Omega-\omega_0)~W_{CR}(x)\\
 &&T^{-1}_{DL}(p) = T^{-1}_{DL} + {\pi\omega_1^2\over\omega_d^2}~(p+1)~g(\Omega - \omega_0)\int dy~{\varphi(y)~y^2\over{p + 1 + \pi\omega_1^2 ~T_{SL}~\varphi(y)}}+\nonumber\\
 &&~~~~~~~+{1\over 2\omega_d^2}\int dx~W_{CR}(x)~x^2\int dy~g^2(y)\label{TDL}\\
 &&\gamma(\omega, p) = {\omega\over\Omega}~p \int_0^\infty d\tau\exp(-p\tau)\left[\beta(\omega, T_{SL}~\tau)/\beta_L - 1\right]\\
 &&\gamma_d(p) = p\int_0^\infty d\tau\exp(-p\tau)\left[\beta_d(T_{SL}~\tau)/\beta_L - 1\right]
\end{eqnarray}
with the equilibrium initial conditions:
\begin{equation}
 \beta(\omega, 0) = \beta_d(0) = \beta_L
\end{equation}
The inverse temperatures are obtained from $\gamma(\omega, p)$ and $\gamma_d(p)$ (which are determined by solving (\ref{gamma1}) and (\ref{gamma2}))\footnote{\, Let us note that in (\ref{gamma1}) the subleading terms containing $\gamma_d(p)$ are omitted; however, those terms do contribute nontrivially to (\ref{gamma2}) via (\ref{TDL}). In the third term on the r.h.s. of (\ref{TDL}) the denominator appearing in the integral in the second term thereof is approximated away due to the narrow width $\Delta$ of the homogeneous line form $\varphi(y)$ (see below). Also, in (\ref{gamma2}) another subleading term is neglected, to wit, that which would stem from the first term in (\ref{gamma1}) due to the aforementioned subleading contribution of $\gamma_d(p)$ into (\ref{gamma1}), which contribution is omitted (see above). Furthermore, in (\ref{gamma1}) and (\ref{TDL}) the function $g(x)$ is treated as constant as its width $\Delta^*$ is much larger than all other relevant quantities, including the characteristic spectral diffusion length in the frequency space (see below). However, in (\ref{gamma2}) the function $g(x)$ is not treated as constant as the integral would vanish in this approximation due to the fact that $\gamma(\omega, p)$ is a symmetric function of $\omega - \Omega$ (see below).} via the inverse Laplace transform.

{}The quantity
\begin{equation}
 \Delta^\prime = [\pi\varphi(0)]^{-1} \left[1 + \pi\omega_1^2~T_{SL}~\varphi(0)\right]^{1/2}
\end{equation}
has the meaning of the frequency interval within which the SP is saturated by the UHF field. If
\begin{equation}\label{add.cond}
 \Delta^\prime \ll \Delta_1
\end{equation}
then the width of the hole burned in the EPR line is determined by the CR and (\ref{gamma1}) can be further approximated as follows:
\begin{eqnarray}
 &&\int d\omega^\prime~f(\omega^\prime - \omega)\left[\gamma(\omega^\prime,p) - \gamma(\omega, p)\right] - (p + 1)~\gamma(\omega,p) = \nonumber\\
 &&~~~~~~~= \pi\omega_1^2~T_{SL}~\varphi(\omega-\Omega)\left[1 + \gamma(\Omega, p)\right]\label{gamma1.1}
\end{eqnarray}
Without delving into this case in detail, let us mention that this equation is integrable in quadratures.

{}In the diffusion approximation $\Delta_{CR} \ll \Delta_1$ \cite{BZK1}, \cite{BZK2}, the nonlocal equation (\ref{gamma1}) can be approximated by a local differential equation:
\begin{equation}
 \Delta_d^2~{\partial^2\gamma(\omega, p)\over\partial\omega^2} - \left[p + 1 + \pi\omega_1^2 ~T_{SL}~\varphi(\omega-\Omega)\right]\gamma(\omega,p) = \pi\omega_1^2~T_{SL}~\varphi(\omega-\Omega)\label{gamma1.d}
\end{equation}
where
\begin{equation}
 \Delta_d^2 = {1\over 2} \int dx~f(x)~x^2 = {\Delta_{CR}^2\over 2}~T_{SL}~g(\Omega - \omega_0) \int dx~W_{CR}(x)
\end{equation}
Now, the order of magnitude of $\Delta_d$ has the meaning of the frequency distance to which spin excitations are propagated within the time $T_{SL}$ due to the CR, which in this case has the characteristics of the so-called spectral diffusion (SD).

{}In the works \cite{BZK1}, \cite{BZK2}, \cite{AD}, \cite{Thesis}, when studying the effect of the SD on the saturation in the stationary case, the additional condition (\ref{add.cond}) was assumed, i.e., it was assumed that $\Delta^\prime \ll \Delta_d$. The results of the instant note (see the next section) hold for general $\Delta^\prime$ and $\Delta_d$ (i.e., here we do {\em not} assume $\Delta^\prime \ll \Delta_d$).

{}Further, usually the function $\varphi(x)$ is assumed to be a truncated Lorentz distribution \cite{KA}. For the reasons which will become clear below, we will assume that
\begin{equation}
 \varphi(x) = (2\Delta)^{-1}\exp(-|x|/\Delta)
\end{equation}
where $\Delta$ has the meaning of the SP width and for inhomogeneous broadening we have $\Delta \ll \Delta^*$. The above approximation is justified as the difference between the exponential form from the Lorentz distribution does not exceed 9\% (while the maximum deviations of the Gaussian from the Lorentz distribution and the exponential form are approximately 23\% and 22\%, respectively).

\section{Solving System of Equations}

{}To determine the inverse temperatures, we must solve the following self-conjugate inhomogeneous boundary problem:
\begin{eqnarray}
 &&{\partial^2\phi(x)\over\partial x^2} - \left[{{p+1}\over\Delta_d^2} + {\pi\omega_1^2\over 2\Delta\Delta_d^2} ~T_{SL}\exp(-|x|/\Delta)\right]\phi(x) = h(x)\\
 &&\lim_{x\rightarrow\pm \infty} \phi(x) = 0
\end{eqnarray}
where the boundary conditions are dictated by the fact that the deviations from the equilibrium occur only in the $\Delta_1$-vicinity of the frequency $\Omega$. We have
\begin{equation}
 \phi(x) = \int dy~ \Gamma(x, y)~h(y)\label{phi}
\end{equation}
where the Green's function $\Gamma(x, y)$ satisfies the following equation
\begin{equation}
 {\partial^2\Gamma(x,y)\over\partial x^2} - \left[{{p+1}\over\Delta_d^2} + {\pi\omega_1^2\over 2\Delta\Delta_d^2} ~T_{SL}\exp(-|x|/\Delta)\right]\Gamma(x,y) = \delta(x-y)\label{Green}
\end{equation}
and the boundary conditions
\begin{equation}
 \lim_{x\rightarrow\pm \infty} \Gamma(x,y) = 0
\end{equation}
Also, the Green's function is symmetric:
\begin{equation}
 \Gamma(x,y) = \Gamma(y,x)\label{sym}
\end{equation}
Via a direct substitution into (\ref{Green}), one can readily verify that
\begin{equation}\label{Gamma}
 \Gamma(x, 0) = -\Delta\left[\lambda~I^\prime_\nu(\lambda)\right]^{-1} I_\nu(\lambda\exp(-|x|/2\Delta))
\end{equation}
where $I_\nu(z)$ is the modified Bessel function of the first kind \cite{Kamke} (also see \cite{PZ}), the prime denotes a derivative w.r.t. the function argument, and
\begin{eqnarray}
 &&\lambda = \omega_1\sqrt{2\pi~T_{SL}~\Delta}/\Delta_d\\
 &&\nu = 2\Delta\sqrt{p+1}/\Delta_d
\end{eqnarray}
Using (\ref{phi}), (\ref{sym}) and (\ref{Gamma}) we can fix $\phi(0)$. Then, for $x\geq 0$ we have the following boundary problem:
\begin{eqnarray}
 &&{\partial^2\phi(x)\over\partial x^2} - \left[{{p+1}\over\Delta_d^2} + {\pi\omega_1^2\over 2\Delta\Delta_d^2} ~T_{SL}\exp(-|x|/\Delta)\right]\phi(x) = h(x)\\
 &&\lim_{x\rightarrow +\infty} \phi(x) = 0\\
 &&\phi(0)~\mbox{is given}
\end{eqnarray}
which can be solved using the fundamental system of solutions of the homogeneous equation (with $h(x) = 0$):
\begin{eqnarray}
 &&\phi_1(x) = I_\nu(\lambda\exp(-x/2\Delta))\\
 &&\phi_2(x) = I_{-\nu}(\lambda\exp(-x/2\Delta))
\end{eqnarray}
Omitting the derivation (which is based on standard techniques), the solution is given by ($x \geq 0$):
\begin{equation}
 \phi(x) = \phi(0)~\phi_1(x)/\phi_1(0) + \int_0^\infty dy~\Gamma^*(x, y)~h(y)
\end{equation}
where (the Heaviside function $\theta(x > 0) = 1$, $\theta(0) = 1/2$, and $\theta(x < 0) = 0$)
\begin{eqnarray}
 &&\Gamma^*(x, y) = {\pi\Delta\over \sin(\pi\nu)}\left\{{I_{-\nu}(\lambda) \over 2I_\nu(\lambda)}~I_\nu\left(\lambda\exp\left(-{x\over 2\Delta}\right)\right) I_\nu\left(\lambda\exp\left(-{y\over 2\Delta}\right)\right)  -  \right.\nonumber\\
 &&~~~~~~~ \left.-\theta(x - y)~I_\nu\left(\lambda\exp\left(-{x\over 2\Delta}\right)\right) I_{-\nu}\left(\lambda\exp\left(-{y\over 2\Delta}\right)\right) + (x\leftrightarrow y)\right\}
\end{eqnarray}
Further, for arbitrary $x$ we have
\begin{equation}
 \phi(x) = \phi(0)~\phi_1(|x|)/\phi_1(0) + \int_0^\infty dy~\Gamma^*(|x|, y) \left[\theta(x)~h(y) + \theta(-x)~h(-y)\right]
\end{equation}
Therefore:
\begin{eqnarray}
 &&\Gamma(x, y) = \theta(x~y)~\Gamma^*(|x|, |y|) - \nonumber\\
 &&~~~~~~~ -\Delta\left[\lambda~I_\nu(\lambda)~I^\prime_\nu(\lambda)\right]^{-1} I_\nu\left(\lambda\exp\left(-{|x|\over 2\Delta}\right)\right) I_\nu\left(\lambda\exp\left(-{|y|\over 2\Delta}\right)\right)
\end{eqnarray}
So, the solution of (\ref{gamma1.d}) is given by
\begin{equation}
 \gamma(\omega, p) = \pi\omega_1^2~T_{SL}~\Delta_d^{-2}\int dy~\Gamma(\omega - \Omega, y)~\varphi(y)
\end{equation}

\section{Limiting Cases}

{}As mentioned above, (\ref{add.cond}) corresponds to the limit where the width of the hole is determined by the SD. This condition can be expressed as follows:
\begin{eqnarray}
 &&\Delta \ll \Delta_d\\
 &&\lambda \ll 1
\end{eqnarray}
which formally is equivalent to taking the limit $\Delta \rightarrow 0$. Using the representation of the function $I_\nu(z)$ via a power series \cite{Baitman}, we have (up to subleading terms in $\Delta$)
\begin{eqnarray}
 &&\Gamma(x,y) = {\Delta_d\over 2\sqrt{p+1}}\left[{\mu\over{\mu + \sqrt{p+1}}}\exp\left(-{\sqrt{p+1}\over\Delta_d}(|x| + |y|)\right)-\right.\nonumber\\
 &&~~~~~~~ \left.-\exp\left(-{\sqrt{p+1}\over\Delta_d}|x - y|\right)\right]
\end{eqnarray}
where
\begin{equation}
 \mu = \pi\omega_1^2~T_{SL} / 2\Delta_d
\end{equation}
is the effective parameter of saturation. For $\gamma(\omega, p)$ and $\gamma_d(p)$ we have
\begin{eqnarray}
 &&\gamma(\omega, p) = - {\mu\over{\mu + \sqrt{p+1}}}\exp\left(-a\sqrt{p+1}\right)\\
 &&\gamma_d(p) = 4\kappa~\Omega~\Delta_d~g^\prime(\Omega - \omega_0)\left[{1\over{\sqrt{p+1}}} - {1\over{\mu + \sqrt{p+1}}}\right]
\end{eqnarray}
and for the inverse temperatures we obtain (via the inverse Laplace transform):
\begin{eqnarray}
 &&[\beta_L - \beta(\omega, t)]/\beta_L = {\mu~\Omega\over\omega}
 \left[- {\mu\over{\mu^2 - 1}}\exp\left(\mu a + (\mu^2 - 1)\tau\right)
 \mbox{erfc}\left({a\over 2\sqrt{\tau}}+\mu\sqrt{\tau}\right) + \right.\nonumber\\
 &&~~~~~~~\left.+ {\exp(-a)\over 2(\mu + 1)}~\mbox{erfc}\left({a\over 2\sqrt{\tau}}- \sqrt{\tau}\right) + {\exp(a)\over 2(\mu - 1)}\times \mbox{erfc}\left({a\over 2\sqrt{\tau}} + \sqrt{\tau}\right) \right]\label{beta1}\\
 &&[\beta_d(t) - \beta_L]/\beta_L = {4\mu\over{\mu^2-1}}~\kappa~\Omega~\Delta_d~g^\prime(\Omega - \omega_0) \times\nonumber\\
 &&~~~~~~~\times\left[\mu~\mbox{erf}(\sqrt{\tau}) - 1 + \exp\left((\mu^2-1)\tau\right)\mbox{erfc}(\mu\sqrt{\tau})\right]
\end{eqnarray}
Here\footnote{\, Taking into account that $T_{SL}/T_{DL} = 2$ or $T_{SL}/T_{DL} = 3$ \cite{Thesis}, typically $\omega_d\ll \Delta_d$, and the first term in (\ref{kappa}) is of order 1, as in \cite{TSU} we can neglect the second term in (\ref{kappa}). Also, the contribution of the second term in (\ref{TDL}) into $\kappa$ in (\ref{kappa}) is negligible in the small $\Delta$ limit. Finally, as in \cite{TSU}, in (\ref{beta1}) we can approximate the overall multiplicative factor $\Omega/\omega$ by 1.}
\begin{eqnarray}
 &&a = |\omega - \Omega| / \Delta_d\\
 &&\tau  = t/T_{SL}\\
 &&\kappa = \left[{\int dy~g^2(y) \over {g(\Omega - \omega_0)}} + {T_{SL}\over{T_{DL}}}~{\omega_d^2 \over \Delta^2_d}\right]^{-1}\label{kappa}
\end{eqnarray}

{}Next, let us consider the limiting case where the contribution of the SD to the saturation of the inhomogeneously broadened line is negligible. This corresponds to taking the limit $\Delta_d\rightarrow 0$, which can be accomplished via the asymptotic formulas for the modified Bessel functions ($z,\nu \rightarrow \infty$) \cite{Baitman} (also see \cite{DLMF}):
\begin{eqnarray}
 &&I_\nu(z) \approx \exp\left((z^2 + \nu^2)^{1/2} - \nu~\mbox{arsh}(\nu/z)\right) / \left[\sqrt{2\pi}~(z^2 + \nu^2)^{1/4}\right]\\
 &&K_\nu(z) \approx \exp\left(\nu~ \mbox{arsh}(\nu/z) - (z^2 + \nu^2)^{1/2}\right) / \left[\sqrt{2/\pi}~(z^2 + \nu^2)^{1/4}\right]
\end{eqnarray}
Here $K_\nu(z)$ is the Macdonald function (the modified Bessel function of the second kind):
\begin{equation}
 I_{-\nu}(z) = I_\nu(z) + 2\sin(\pi\nu)~ K_\nu(z)/\pi 
\end{equation} 
Taking into account the properties of the $\delta$-function
\begin{eqnarray}
 &&\lim_{z\rightarrow +\infty} z~\exp(-z|x|) / 2 = \delta(x)\\
 &&\delta(F(x)) = \delta(x - x_0) / |F^\prime(x_0)|
\end{eqnarray}
where $F(x_0)=0$, we get (up to subleading terms in $\Delta_d$)
\begin{equation}
 \Gamma(x,y) = -\Delta_d^2\left[p+1+\pi\omega_1^2~T_{SL}~\varphi(x)\right]^{-1}\delta(x-y)
\end{equation}
as it should be.

\section{Concluding Remarks}

{}The choice of the function $\varphi(x)$ as the exponential line form is motivated by the fact that (\ref{gamma1.d}) in this case can be solved in quadratures via modified Bessel functions, which are well-studied. On the other hand, when $\varphi(x)$ has the form of the (truncated) Lorentz distribution, the Green's function $\Gamma(x,y)$ cannot be expressed via elementary or known special functions.

\end{document}